\documentclass[aps,pra,10pt,twocolumn,superscriptaddress,showkeys,letterpaper]{revtex4-2}

\pdfoutput=1

\usepackage[T1]{fontenc}
\usepackage[utf8]{inputenc}
\usepackage{color}
\usepackage{babel}
\usepackage{mathtools}
\usepackage{amsmath}
\usepackage{amsthm}
\usepackage{amssymb}

\usepackage{graphicx}
\usepackage[pdfusetitle,
 bookmarks=true,bookmarksnumbered=false,bookmarksopen=false,
 breaklinks=false,pdfborder={0 0 0},pdfborderstyle={},backref=false,colorlinks=true]
 {hyperref}
\hypersetup{
 linkcolor=magenta, urlcolor=blue, citecolor=blue}

\makeatletter

\usepackage{caption}

\usepackage{subcaption}

\DeclareMathOperator{\Tr}{Tr}

\DeclareMathOperator{\sech}{sech}

\allowdisplaybreaks

\makeatother

\begin{document}
\title{Canonical quantization of neurons}
\author{Alexander He}
\affiliation{Department of Physics, Cornell University, Ithaca, New York 14850, USA}
\author{Nana Liu}
\affiliation{Institute of Natural Sciences, School of Mathematical Sciences, Ministry of Education Key Laboratory in Scientific and Engineering Computing, and Global College, Shanghai Jiao Tong University, Shanghai 200240, China}
\author{Mark M. Wilde}
\affiliation{School of Electrical and Computer Engineering,
Cornell University, Ithaca, New York 14850, USA}

\begin{abstract}
Canonical quantization provides a systematic procedure for constructing
quantum models from classical Hamiltonians. Here, we apply this principle to a fundamental
computational primitive of machine learning: the neuron. Specifically,
by viewing a neuron as a composition of an energy function and an
activation function, we quantize this model by replacing the energy
function with a quantum Hamiltonian and applying the activation function
to it through matrix functional calculus. This results in an activation observable that can be measured
on an input quantum state. We investigate the use of these quantized
neurons for function approximation, where the objective is to learn
an unknown observable from labeled quantum data. For this purpose,
we develop hybrid quantum--classical algorithms for training and evaluation, including
procedures for measuring the activation observable and estimating gradients of the squared loss error. Our
algorithms for gradient estimation rely on basic primitives like classical
random sampling, the Hadamard test, and Hamiltonian simulation, and
those for measuring an activation observable rely on quantum algorithms known as the power of
one qumode and Schr\"odingerization. Numerical experiments demonstrate
that our quantized neurons exhibit enhanced expressive capabilities
relative to corresponding classical neurons on representative learning
tasks. Our work establishes canonical quantization as a principled
framework for constructing quantum machine learning primitives and
provides a foundation for developing neural architectures tailored
to quantum data.
\end{abstract}

\maketitle

\textit{Introduction}---Quantum mechanics furnishes a foundational
framework for describing physical systems across atomic, molecular,
optical, condensed matter, and high-energy physics~\cite{cohen2020quantum1,cohen2020quantum2,cohen2020quantum3}.
Over the past century, its principles have been extended to new domains
through the emergence of quantum information science, enabling unconventional
approaches to computation~\cite{Dalzell2025} and communication~\cite{Gisin2007}.
During the past decade, quantum information science has motivated
efforts to understand whether intrinsically quantum effects can enhance machine learning~\cite{chang2025primer}.

A central principle underlying the development of quantum mechanics
is canonical quantization~\cite{Shankar1994}, which promotes classical
variables to operators acting on a Hilbert space. For example, a classical
Hamiltonian $H(x,p)$ is transformed into a quantum Hamiltonian $H(\hat{x},\hat{p})$,
where the canonical position and momentum variables $x$ and $p$,
respectively, are promoted to noncommuting operators satisfying $\left[\hat{x},\hat{p}\right]=i\hbar$.
This procedure provides a systematic route for constructing quantum
models from classical descriptions and has played a foundational role 
throughout modern physics. 

Here, we introduce a different application of canonical quantization:
rather than quantizing physical dynamical systems, we quantize a neuron,
a fundamental computational primitive of machine learning.
To the best of our knowledge, no prior construction of a quantum neuron
has followed the canonical quantization procedure while also demonstrating
an effective training and evaluation protocol. Existing
approaches to quantum neurons typically encode neural information
into quantum states or amplitudes~\cite{Kapoor2016,cao2017,Wan2017,Hu2018QuantumNeuron,Killoran2019,Yan2020,Kristensen2021,Monteiro2021,Singh2024,barney2025,Roncallo2025},
rather than constructing a quantum observable by applying canonical
quantization to the underlying classical neuron model. Consequently,
they do not directly realize the canonical map from classical variables
to quantum operators that underlies quantum mechanics. 

Prior to our proposal here, the concept of canonical quantization
has been successfully applied to a different classical machine learning
model known as Boltzmann machines~\cite{Ackley1985,Hinton1986}. To
gain a glimpse of how quantum Boltzmann machines~\cite{Amin2018,Benedetti2017,Kieferova2017}
are constructed, let us first briefly recall classical Boltzmann machines.
A classical Boltzmann machine models data through a thermal distribution
$q_{\theta}(v,h)\propto e^{-E_{\theta}(v,h)}$, where $E_{\theta}(v,h)\coloneqq\sum_{j\in [J]}\theta_{j}E_{j}(h,v)$
is a classical Hamiltonian parameterized by the parameter vector $\theta\coloneqq(\theta_{1},\ldots,\theta_{J})$,
and each $E_{j}(h,v)$ is a fixed energy interaction term. Canonical
quantization replaces the energy function $E_{\theta}(v,h)$
with a Hamiltonian operator
\begin{equation}
H_{\theta}\coloneqq\sum_{j\in [J]}\theta_{j}H_{j},\label{eq:gen-param-Ham}
\end{equation}
resulting in the quantum thermal state $\sigma_{\theta}\propto e^{-H_{\theta}}$.
When the terms in $H_{\theta}$ do not commute, the resulting
model has no classical probabilistic description.

Inspired by the canonical quantization of Boltzmann machines, we apply
the same principle to the fundamental computational primitive of machine
learning: the neuron. 
Specifically, we interpret the energy function underlying a classical neuron as a parameterized Hamiltonian,
and we quantize this Hamiltonian by replacing its classical variables
with noncommuting operators, yielding $H_{\theta}$. The activation
function $\varphi$ is then promoted to a matrix function of this Hamiltonian,
producing an activation observable~$\varphi(H_{\theta})$. For example,
as identified in~\cite{liu2026fermidirac}, choosing $\varphi(H_{\theta})=(e^{-H_{\theta}/T}+I)^{-1}$,
with $T>0$ a temperature parameter, gives a quantum generalization
of the sigmoid function used in classical neurons. Whereas quantum
Boltzmann machines encode learning models in quantum states through
the Gibbs map $H_{\theta}\mapsto e^{-H_{\theta}}/\Tr[e^{-H_{\theta}}]$,
our approach encodes neural transformations directly as quantum observables.
Thus, quantized neurons provide a complementary paradigm in which
the nonlinear processing element of machine learning, rather than
the underlying probability distribution, is quantized.

Key contributions of our paper include our approach
for quantizing neurons, hybrid quantum--classical (HQC) algorithms
for evaluating and training them, and numerical experiments demonstrating
learning tasks in which quantized neurons exhibit behavior outperforming corresponding classical models on representative learning tasks. Interestingly, our HQC algorithms
for training make use of basic primitives that include classical random
sampling, the Hadamard test~\cite{Cleve1998}, and Hamiltonian simulation~\cite{lloyd1996universal,childs2018toward}, similar to
recent HQC algorithms proposed for training quantum Boltzmann machines~\cite{Patel2025a,Patel2025}. Here we focus on neurons
built from the hyperbolic tangent activation function, while our companion
paper~\cite{he2026} develops the quantization procedure for five additional activation functions that play an essential role in 
machine learning. This includes the smooth rectified linear unit (ReLU), sigmoid linear unit, Gaussian error function, Gaussian-smoothed ReLU, and Gaussian error linear unit. 

\textit{Canonical quantization of neurons}---Let us begin by recalling
the perceptron model~\cite{Rosenblatt1958Perceptron} of a classical
neuron. For $n\in\mathbb{N}$, suppose that $z\coloneqq\left(z_{1},\ldots,z_{n}\right)\in\left\{ -1,1\right\} ^{n}$
is a vector of input spin variables; in our companion paper~\cite[Sec.~V]{he2026},
we consider the more general case when each $z_{i}\in\mathbb{R}$. A classical
neuron consists of a composition of an energy function $w^{T}z+b$
and an activation function $\varphi$:
\begin{equation}
\varphi(w^{T}z+b),\label{eq:classical-neuron}
\end{equation}
where $w\in\mathbb{R}^{n}$ is a weight vector and $b\in\mathbb{R}$
is a bias. By training it on input data, the weight vector and bias
can be tuned for tasks like function approximation or classification.

Before quantizing the neuron model in~\eqref{eq:classical-neuron},
let us first rewrite the energy function as follows:
$
w^{T}z+b=\sum_{i=1}^{n}w_{i}z_{i}+b$.
Now following canonical quantization, we promote each classical variable
$z_{i}$ to an operator~$\hat{z}_{i}$. Given that each input variable
$z_{i}$ is a classical spin variable taking values in $\left\{ -1,1\right\} $ by assumption, each
operator $\hat{z}_{i}$ should be Hermitian with eigenvalues $-1$
and $+1$. As such, a natural choice for quantizing $z_{i}$ is to
set $\hat{z}_{i}=\sigma_{Z}^{(i)}$, where $\sigma_{Z}^{(i)}$ is
the Pauli $Z$ matrix acting on the $i$th qubit (i.e., $\sigma_{Z}^{(i)}\coloneqq I^{\otimes i-1}\otimes\sigma_{Z}\otimes I^{\otimes n-i}$).
In doing so, the classical energy function above
is promoted to the Hamiltonian
$
H_{C}(\theta)\coloneqq\sum_{i=1}^{n}w_{i}\sigma_{Z}^{(i)}+bI^{\otimes n}$,
where $\theta\equiv\left(w,b\right)$. Now applying the activation
function $\varphi$ to $H_{C}(\theta)$, we arrive at the 
activation observable
$
\varphi(H_{C}(\theta))$.
This observable can be measured on an input state $\rho$, leading
to a measurement outcome with expected value $\Tr\!\left[\varphi(H_{C}(\theta))\rho\right]$.

Let $|z\rangle\equiv|z_{1}\rangle\otimes\cdots\otimes|z_{n}\rangle$
be a tensor product of $\pm1$-eigenstates of $\sigma_{Z}$. One finds
that $H_{C}(\theta)|z\rangle=\left(w^{T}z+b\right)|z\rangle$, so
that $|z\rangle$ is an eigenvector of $H_{C}(\theta)$ with corresponding
eigenvalue $w^{T}z+b$. By applying the functional calculus, we then
conclude that the activation observable $\varphi(H_{C}(\theta))$
admits the following spectral decomposition:
\begin{equation}
\varphi(H_{C}(\theta))=\sum_{z\in\left\{ -1,1\right\} ^{n}}\varphi(w^{T}z+b)|z\rangle\!\langle z|.\label{eq:classical-act-obs}
\end{equation}
Evaluating the expected value of $\varphi(H_{C}(\theta))$ with respect
to an input state $\rho$ then leads to
$
\Tr\!\left[\varphi(H_{C}(\theta))\rho\right]=\sum_{z\in\left\{ -1,1\right\} ^{n}}p(z)\,\varphi(w^{T}z+b)$,
where $p(z)\equiv\langle z|\rho|z\rangle$ is the probability distribution
that results from measuring~$\rho$ in the orthonormal basis $\left\{ |z\rangle\right\} _{z\in\left\{ -1,1\right\} ^{n}}$.
As such, our quantization of~\eqref{eq:classical-neuron} reduces
exactly to the classical case, a consistency requirement that should hold
for any plausible quantization of a neuron. Indeed, one can simulate
the measurement of $\varphi(H_{C}(\theta))$ simply by measuring the
input state $\rho$ in the basis $\left\{ |z\rangle\right\} _{z\in\left\{ -1,1\right\} ^{n}}$
and feeding the outcome $z$ into $\varphi$.

We can generalize the approach above to allow for non-classical neurons
by taking the underlying Hamiltonian to be a quantum Hamiltonian of
the following form:
\begin{equation}
\sum_{\alpha,\beta,i,j }\Omega_{\left(\alpha,i\right),\left(\beta,j\right)}\sigma_{\alpha}^{\left(i\right)}\otimes\sigma_{\beta}^{\left(j\right)}
+\sum_{\alpha,i }\omega_{\alpha,i}\sigma_{\alpha}^{\left(i\right)}+bI^{\otimes n},\label{eq:fully-q-ham}
\end{equation}
where the sums over $\alpha,\beta$ are over $\{x,y,z\}$ and those over
$i,j$ are over $[n]$. Also, $\theta\coloneqq\left(\Omega,\omega,b\right)$, $\Omega\in\mathbb{R}^{3n\times3n}$,
$\omega\in\mathbb{R}^{3n}$, and the interacting and noninteracting
terms include all Pauli-$X$, $Y$, and $Z$ observables indexed by
$x$, $y$, and $z$. The Hamiltonian in~\eqref{eq:fully-q-ham} includes
non-commuting and interaction terms, hallmarks of Hamiltonians like
the transverse-field Ising~\cite[Chap.~5]{Sachdev2011} and Heisenberg
models~\cite{Mattis2006,Goldschmidt2011} that play key roles in quantum
physics.

We can take an even broader approach by allowing for all
Hamiltonians of the form in~\eqref{eq:gen-param-Ham}, where $\theta_{j}\in\mathbb{R}$
and $H_{j}$ is a Hamiltonian operator acting nontrivially on $k$
of the $n$ qubits, where $k$ is a constant independent of $n$.
A distinguishing feature of such quantum Hamiltonians is that they are
generally computationally difficult to diagonalize classically;
however, quantum algorithms can access spectral information
through procedures related to the quantum phase estimation algorithm~\cite{kitaev1995}. In this broader approach, an activation observable is given by $\varphi(H(\theta))$, and one can measure this observable on an input state $\rho$.

\textit{Representative quantized neuron}---A concrete example, which we focus on here,
consists of setting $\varphi$ to be the temperature-scaled hyperbolic tangent
activation function, abbreviated as $g_{T}(x)\coloneqq\tanh(x/T)$,
where $T>0$ is a temperature parameter that controls the sharpness
of the threshold function. In doing so, we arrive at the activation
observable $g_{T}(H(\theta))$, where $H(\theta)$ is a parameterized
Hamiltonian of the form in~\eqref{eq:gen-param-Ham}.

\textit{Function approximation}---We now formulate a learning task for which quantized
neurons provide a natural hypothesis class: function approximation. Before introducing
the quantum version of this problem, let us first recall the classical
problem that involves a single neuron. Here, we suppose that there
is an unknown function $f\colon\left\{ -1,1\right\} ^{n}\to\left[-1,1\right]$
to learn, while training data is available in the form $\left(z^{(1)},y_{1}\right),\ldots,\left(z^{(M)},y_{M}\right)$,
where $M\in\mathbb{N}$ and $z^{\left(m\right)}\in\left\{ -1,1\right\} ^{n}$,
$y_{m}\in\left[-1,1\right]$ for all $m\in\left[M\right]$. Each pair
satisfies $f(z^{\left(m\right)})=y_{m}$, and the goal is to learn
this mapping. One can then use $g_{T}(w^{T}z+b)$ as an approximation
of $f(z)$ and employ the training data to minimize the mean squared
error:
\begin{equation}
\frac{1}{M}\sum_{m=1}^{M}\left(g_{T}(w^{T}z^{\left(m\right)}+b)-y_{m}\right)^{2}.\label{eq:loss-function}
\end{equation}
The weight vector $w$ and bias $b$ are trained by performing gradient
descent on the loss function in~\eqref{eq:loss-function}, which requires
access to the gradient of~\eqref{eq:loss-function}.

We can embed this classical learning problem into a quantum learning
problem by encoding each $z^{\left(m\right)}$ into a quantum state
$\rho_{m}=|z^{\left(m\right)}\rangle\!\langle z^{\left(m\right)}|$
and the function $f$ into an observable $O=\sum_{z\in\left\{ -1,1\right\} ^{n}}f(z)|z\rangle\!\langle z|$.
Then the activation observable in~\eqref{eq:classical-act-obs} (with
$\varphi$ therein set to $g_{T}$) can be used as an approximation
of $O$ in the  loss function
$
\frac{1}{M}\sum_{m=1}^{M}\left(\Tr\!\left[g_{T}(H_{C}(\theta))\rho_{m}\right]-y_{m}\right)^{2}$,
which reduces exactly  to the classical case in~\eqref{eq:loss-function}
 because $\Tr\!\left[g_{T}(H_{C}(\theta))\rho_{m}\right]=g_{T}(w^{T}z^{\left(m\right)}+b)$
for our choices above.

The above construction motivates a general quantum function approximation problem that
goes beyond the classical case. Here, the goal is to approximate an
unknown observable $O$ through its expectation values on a set of input states. We suppose that quantum
training data is available in the form $\left(\rho_{1},y_{1}\right),\ldots,\left(\rho_{M},y_{M}\right)$,
where $\rho_{m}$ is a quantum state and $y_{m}\in\left[-1,1\right]$
for all $m\in\left[M\right]$. Each pair satisfies $\Tr\!\left[O\rho_{m}\right]=y_{m}$,
and the goal is to learn this mapping. We can then pick a general
parameterized Hamiltonian $H(\theta)$ of the form in~\eqref{eq:gen-param-Ham}
and use $g_{T}(H(\theta))$ as an approximation of $O$, with the
goal being to minimize the loss function
$
\mathcal{L}(\theta)\coloneqq\frac{1}{M}\sum_{m=1}^{M}\left(\Tr\!\left[g_{T}(H(\theta))\rho_{m}\right]-y_{m}\right)^{2}$.
Employing the shorthand $\partial_{j}\equiv\frac{\partial}{\partial\theta_{j}}$,
the $j$th element $\partial_{j}\mathcal{L}(\theta)$ of the gradient $\nabla_{\theta}\mathcal{L}(\theta)$
is given by
\begin{equation*}
\frac{2}{M}\sum_{m=1}^{M}\left(\Tr\!\left[g_{T}(H(\theta))\rho_{m}\right]-y_{m}\right) \partial_{j}\Tr\!\left[g_{T}(H(\theta))\rho_{m}\right].
\end{equation*}
Setting $s\in\mathbb{N}$ and $\eta>0$ a hyperparameter known as
the learning rate, we can iterate the gradient-descent update rule
$\theta^{(s+1)}\leftarrow\theta^{\left(s\right)}-\eta\left.\nabla_{\theta}\mathcal{L}(\theta)\right|_{\theta=\theta^{\left(s\right)}}$ in order
to find a local minimum of $\mathcal{L}(\theta)$. 

Suppose now that $\rho_{m}$ is a quantum state to which we have sample
access and $H(\theta)$ is a parameterized local Hamiltonian. Then
evaluating $\Tr\!\left[g_{T}(H(\theta))\rho_{m}\right]$ and $\partial_{j}\Tr\!\left[g_{T}(H(\theta))\rho_{m}\right]$
is generally computationally intractable for classical methods, and one can instead
employ a quantum computer to estimate these quantities, as delineated
in what follows.

\begin{figure}

\begin{centering}
\includegraphics[width=3.2in]{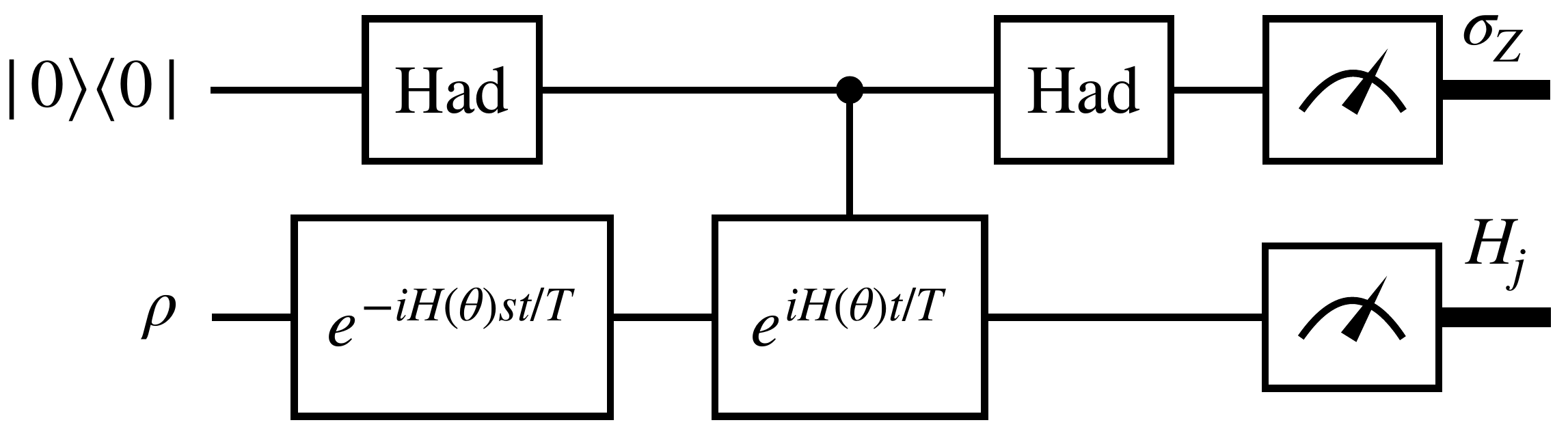}\caption{Quantum circuit for estimating $\partial_{j}\Tr\!\left[g_{T}(H(\theta))\rho\right]$.}\label{fig:had-test-grad}
\par\end{centering}
\end{figure}

\textit{Hybrid quantum--classical algorithms for gradient estimation}---Given
the application mentioned above, it is important to have a method
for estimating $\partial_{j}\Tr\!\left[g_{T}(H(\theta))\rho\right]$
for all $j\in\left[J\right]$ and for every state~$\rho$. Consider
the following integral representation:
$
g_{T}(x)=\mathbb{E}_{t\sim\mu}\!\left[\frac{e^{itx/T}}{it}\right],
$
where $\mu(t)\coloneqq\frac{t}{2\sinh(\pi t/2)}$ is a probability
density function. This representation expresses the nonlinear activation
function $g_T$ as an average over phases. By employing it and Duhamel's
formula for the derivative of the matrix exponential, we find that
\begin{align}
 & \partial_{j}\Tr\!\left[g_{T}(H(\theta))\rho\right]\nonumber \\
 & =\partial_{j}\Tr\!\left[\mathbb{E}_{t\sim\mu}\!\left[\frac{e^{itH(\theta)/T}}{it}\right]\rho\right]\nonumber \\
 & =\mathbb{E}_{t\sim\mu}\!\left[\frac{1}{it}\Tr\!\left[\partial_{j}e^{itH(\theta)/T}\rho\right]\right]\nonumber \\
 & =\mathbb{E}_{t\sim\mu}\!\left[\frac{1}{it}\Tr\!\left[\mathbb{E}_{s\sim\upsilon}\!\left[e^{istH(\theta)/T}\frac{itH_{j}}{T}e^{i\left(1-s\right)tH(\theta)/T}\right]\rho\right]\right]\nonumber \\
 & =\frac{1}{T}\mathbb{E}_{t\sim\mu,s\sim\upsilon}\!\left[\Tr\!\left[H_{j}e^{itH(\theta)/T}\mathcal{U}_{st/T}^{H(\theta)}\left(\rho\right)\right]\right],\label{eq:grad-estimate}
\end{align}
where $\upsilon$ is a uniform probability density over the unit interval
$\left[0,1\right]$ and $\mathcal{U}_{st/T}^{H(\theta)}(\rho)\coloneqq e^{-istH(\theta)/T}\rho e^{istH(\theta)/T}$
is a unitary quantum channel. Recognizing that~\eqref{eq:grad-estimate}
takes a form similar to what the standard Hadamard test quantum circuit
estimates, we can repeatedly perform the following procedure and calculate
a sample average at the end, in order to estimate $\partial_{j}\Tr\!\left[g_{T}(H(\theta))\rho\right]$:
1)~sample $t\sim\mu$ and $s\sim\upsilon$, 2)~execute the Hadamard
test quantum circuit depicted in Fig.~\ref{fig:had-test-grad},
and 3)~multiply the measurement outcomes to obtain a random outcome
with expected value equal to the last line in~\eqref{eq:grad-estimate}.
By the law of large numbers (in particular, the Hoeffding inequality~\cite{Hoeffding1963}),
the resulting sample average is an unbiased
estimate of~\eqref{eq:grad-estimate} that converges quickly to it (see~\cite[Eq.~(29)]{he2026}).

\textit{Quantum algorithm for measuring an activation observable}---The
other key ingredient needed for training and testing is a quantum
algorithm for measuring the activation observable $g_{T}(H(\theta))$.
Since we have the gradient expression in~\eqref{eq:grad-estimate},
we can integrate it and arrive at an expression for $\Tr\!\left[g_{T}(H(\theta))\rho\right]$
that can be estimated by means of a procedure similar to that outlined
above (see~\cite[Alg.~2]{he2026}).

\begin{figure}
\begin{centering}
\includegraphics[width=3in]{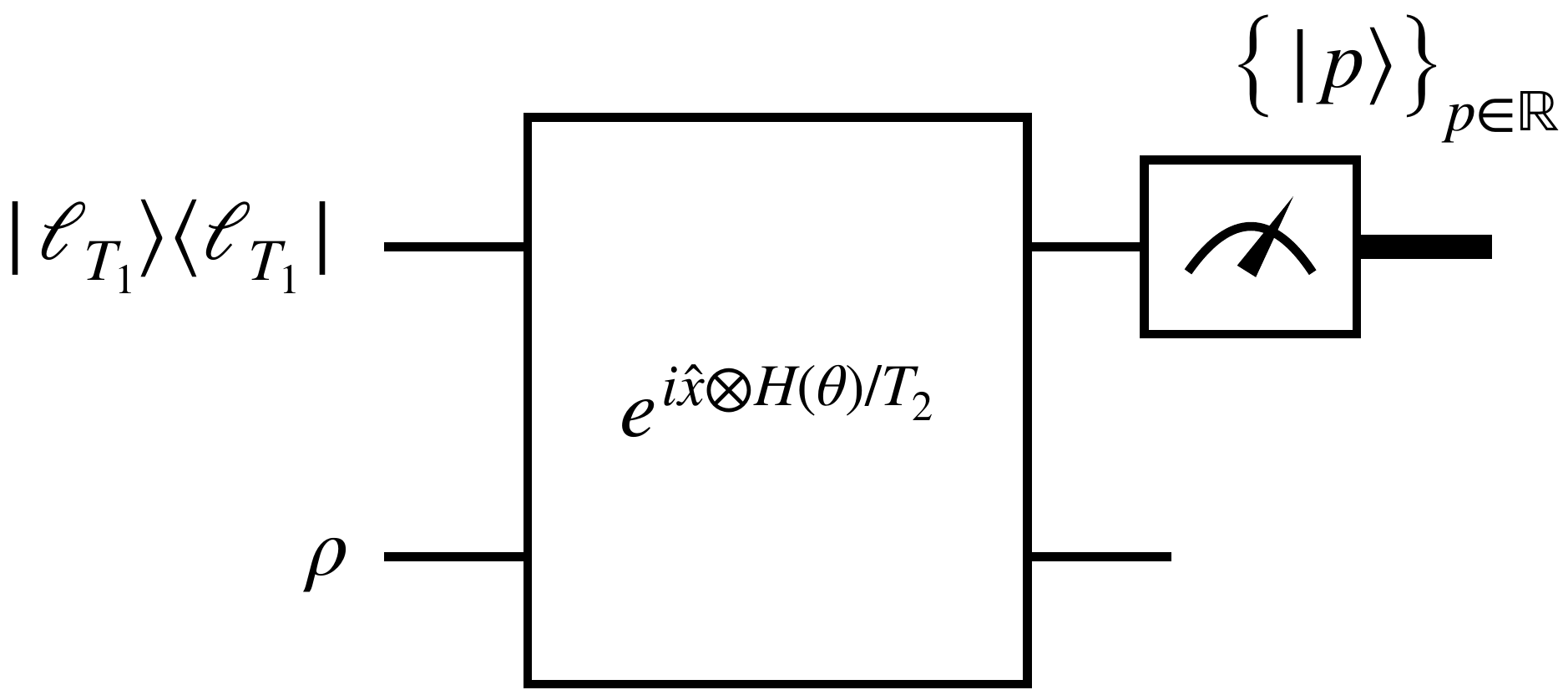}\caption{Circuit for measuring  $g_{T}(H(\theta))$
on input  $\rho$.}\label{fig:q-alg-meas-act-obs}
\par\end{centering}
\end{figure}

Alternatively, we can employ a method that builds on primitives known
as the power of one qumode~\cite{Liu2016} and Schr\"odingerization~\cite{Jin2023,Jin2024}.
Here we employ a control qumode
that interacts with the $n$ qubits of the quantized neuron. The method begins
by preparing a control qumode in the state
$
|\ell_{T_{1}}\rangle\coloneqq\int_{\mathbb{R}}dp\,\sqrt{\ell_{T_{1}}(p)}|p\rangle$,
where $\ell_{T_{1}}$ is the following logistic probability density
function:
\begin{equation}
\ell_{T_{1}}(p)\coloneqq\frac{e^{p/T_{1}}}{T_{1}\left(e^{p/T_{1}}+1\right)^{2}}=\frac{1}{4T_{1}}\sech^{2}\!\left(\frac{p}{2T_{1}}\right),\label{eq:logistic-prob-dens}
\end{equation}
$\left\{ |p\rangle\right\} _{p\in\mathbb{R}}$ denotes the momentum
quadrature eigenbasis, and $T_{1},T_{2}>0$ are chosen such that $T/2=T_{1}T_{2}$.
Then, prepare a data register of $n$ qubits in the state $\rho$,
interact the control qumode and the data register according to the
Hamiltonian evolution $e^{i\hat{x}\otimes H(\theta)/T_{2}}$, and measure
the momentum quadrature of the control qumode to obtain the outcome
$p\in\mathbb{R}$. Finally, set the output $Y=1$ if $p\geq0$ and
$Y=-1$ if $p<0$. The expected value of $Y$
is equal to $\Tr\!\left[g_{T}(H(\theta))\rho\right]$, as stated in~\cite[Thm.~4]{he2026}, thus providing an effective procedure for
measuring $g_{T}(H(\theta))$. Fig.~\ref{fig:q-alg-meas-act-obs} depicts this quantum
algorithm.

\textit{Numerical experiments}---We conducted numerical experiments
to compare the performance of our quantized neurons against classical
neurons for the task of function approximation.
We considered a realizable setting where the target observable $O$ is generated by a transverse-field Ising Hamiltonian passed through the nonlinear activation $g_T$. That is,
we took the unknown
observable $O$ to be $g_{T}(H^{\star}(\zeta))$, where $H^{\star}(\zeta)$
is a transverse-field Ising model and $\zeta$ is a randomly selected
parameter vector. We generated training data by picking each state
$\rho_{m}$ to be a tensor product of Pauli $X$, $Y$, and $Z$ eigenstates,
Bell states, and maximally mixed states. We generated each $y_{m}$
by calculating $y_{m}=\Tr\!\left[g_{T}(H^{\star}(\zeta))\rho_{m}\right]$.
We took the quantum model to be a transverse-field Ising model and
the classical model to be a classical Ising model, defined as follows:
\begin{align*}
H_{\text{TFIM}}(\theta) & \coloneqq\sum_{i=1}^{n-1}W_{i}\sigma_{Z}^{\left(i\right)}\otimes\sigma_{Z}^{\left(i+1\right)}+\sum_{i=1}^{n}w_{i}\sigma_{X}^{\left(i\right)}+bI^{\otimes n},\\
H_{\text{IM}}(\theta) & \coloneqq\sum_{i=1}^{n-1}W_{i}\sigma_{Z}^{\left(i\right)}\otimes\sigma_{Z}^{\left(i+1\right)}+\sum_{i=1}^{n}w_{i}\sigma_{Z}^{\left(i\right)}+bI^{\otimes n},
\end{align*}
with $\theta=\left(W,w\right)$ and 
the key difference between TFIM and IM being that the TFIM is noncommuting due to the transverse $X$-field, whereas the Ising model is fully commuting in the computational basis. Observe that
the number of parameters is the same for each model.

\begin{figure}
\centering


\begin{centering}
\includegraphics[width=3in]{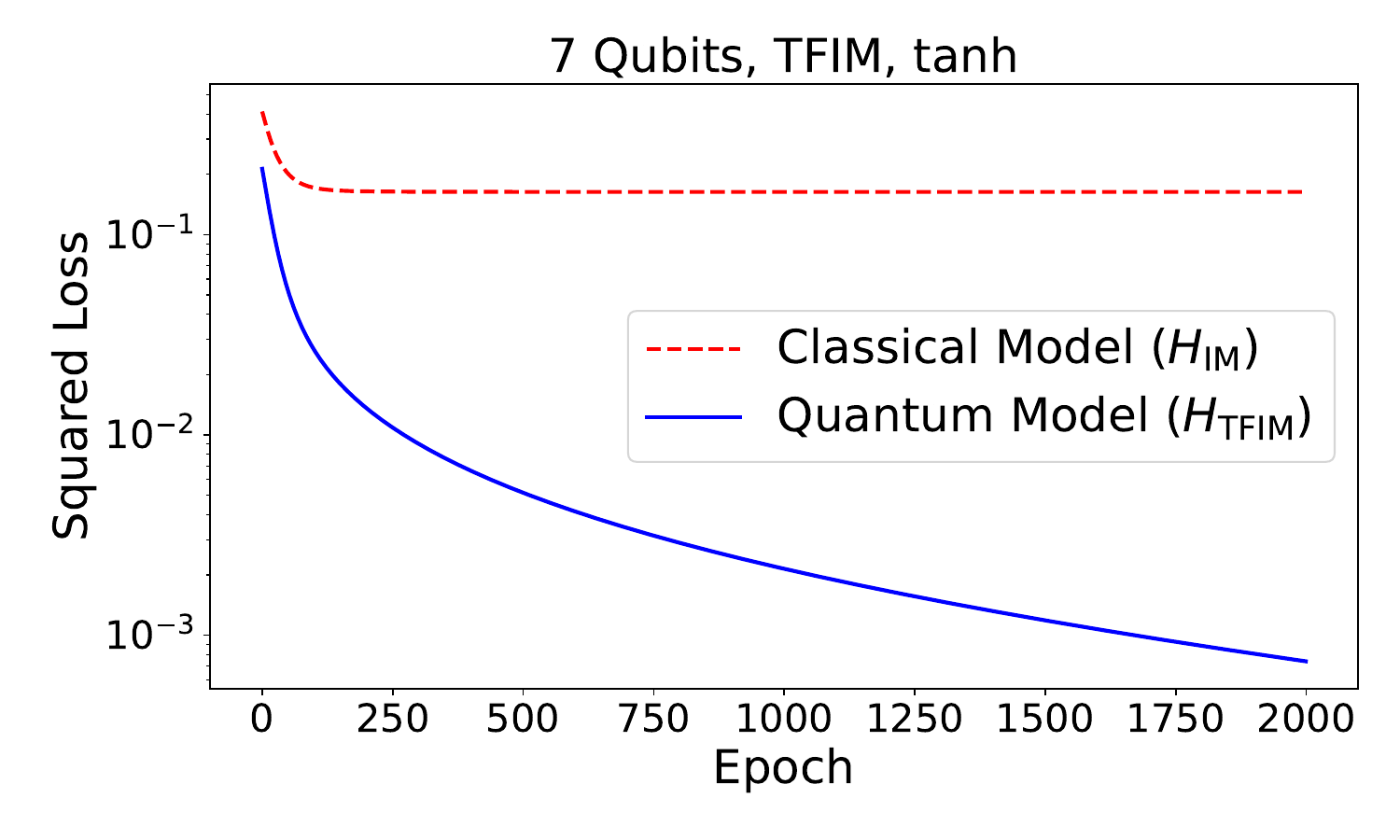}
\end{centering}
\caption{Training performance for squared-loss minimization using
quantized neurons with transverse-field Ising (TFIM) and commuting
Ising (IM) Hamiltonians, and hyperbolic tangent activation, for
a seven-qubit system.}
\label{fig:numerics}
\end{figure}

Fig.~\ref{fig:numerics} depicts training performance for squared
loss minimization. The experiment included learning using $\tanh$
for a seven-qubit system. In this experiment, the noncommuting
Hamiltonian model outperformed the commuting model when
learning observables generated by nonlinear functions of Hamiltonians
(see \cite[Sec.~VI]{he2026} for more experiments consistent with these
findings). This suggests that the expressive power of the underlying
operator algebra plays a central role in the learnability of quantum-induced
functions in this setting. Intuitively, this enhancement arises because
noncommuting Hamiltonians generate additional directions in operator space that are
inaccessible to commuting models. For example, while $H_{\operatorname{TFIM}}$
contains $\sigma_X$ terms that generate sensitivity to $X$-basis structure,
$H_{\operatorname{IM}}$ is diagonal in the computational basis and
cannot directly represent such contributions. Furthermore, the
activation observable $g_T(H_{\text{TFIM}}(\theta))$
contains higher-order Pauli terms, including terms proportional to $\sigma_Y$.
These arise because the Taylor expansion of $g_T$ generates powers of
$H_{\text{TFIM}}(\theta)$, and products of noncommuting Pauli operators such
as $\sigma_X\sigma_Z=i\sigma_Y$ produce even more operator directions that are
absent in the commuting Ising model.

\textit{Conclusion}---In summary, we introduced a method for quantizing
classical neurons based on canonical quantization, in which nonlinear
activation functions are applied to parameterized quantum Hamiltonians
in order to define activation observables.
We developed hybrid quantum--classical algorithms that can be used
to train and test these quantized neurons on quantum data. While focusing
on the hyperbolic tangent function here, our companion paper~\cite{he2026}
develops these same tools for five other activation functions that
have played a prominent role in classical machine learning. We showcased
the results of simple numerical experiments demonstrating that
our quantized neurons feature enhanced expressive power relative to
corresponding classical neurons.

Our companion paper~\cite{he2026} includes several
results not presented here, including logistic loss minimization for
binary classification of quantum data, extensions to quantized neurons
for qumodes and continuous-variable quantum data, a BQP-complete computational
decision problem based on our quantized neurons, and, finally, two
methods for composing these quantized neurons into networks. One practical
method for forming networks involves a first layer of quantized neurons
for processing quantum data, while subsequent layers are purely classical.

Several directions remain open for future work. An important open question
is whether quantized neurons exhibit barren-plateau behavior~\cite{Larocca2025}.
Existing analyses of barren plateaus, which primarily focus on parameterized
quantum circuits and related variational quantum algorithms, do not directly
apply to this setting, motivating a separate investigation of trainability. 
Next, extending numerical simulations to larger system sizes and experimentally
relevant quantum data sets is an important future direction. Finally,
it would be illuminating to investigate the aforementioned hybrid neural network architectures in
which the first layer consists of quantized neurons, while subsequent layers
are classical. Such architectures combine quantum processing of quantum data
with the flexibility of classical machine learning methods.

\begin{acknowledgments}
We thank Zo\"e Holmes and her research group members for feedback.
NL acknowledges funding from the Science
and Technology Commission of Shanghai Municipality (STCSM) grant
no.~24LZ1401200 (21JC1402900), NSFC grants no.~12471411 and no.~12341104,
the Shanghai Jiao Tong University 2030 Initiative, the Shanghai Pilot
Program for Basic Research, and the Fundamental Research Funds for
the Central Universities. MMW acknowledges support from the National
Science Foundation under grant no.~2329662.
\end{acknowledgments}

\bibliography{Ref}
 
\end{document}